# Dataflow-Aware PIM-Enabled Manycore Architecture for Deep Learning Workloads

Harsh Sharma[1], *Student Member, IEEE*, Gaurav Narang[1], *Student Member, IEEE,* Janardhan Rao Doppa[1], *Senior Member, IEEE,* Umit Ogras[2], *Senior Member, IEEE,* and Partha Pratim Pande[1], *Fellow, IEEE*
[1]*Washington State University Pullman, WA;* [2]*University of Wisconsin Madison, Madison, WI*

*Abstract*— Processing-in-memory (PIM) has emerged as an enabler for the energy-efficient and high-performance acceleration of deep learning (DL) workloads. Resistive random-access memory (ReRAM) is one of the most promising technologies to implement PIM. However, as the complexity of Deep convolutional neural networks (DNNs) grows, we need to design a manycore architecture with multiple ReRAM-based processing elements (PEs) on a single chip. Existing PIM-based architectures mostly focus on computation while ignoring the role of communication. ReRAM-based tiled manycore architectures often involve many Processing Elements (PEs), which need to be interconnected via an efficient on-chip communication infrastructure. Simply allocating more resources (ReRAMs) to speed up only computation is ineffective if the communication infrastructure cannot keep up with it. In this paper, we highlight the design principles of a dataflow-aware PIM-enabled manycore platform tailor-made for various types of DL workloads. We consider the design challenges with both 2.5D interposer- and 3D integration-enabled architectures.

## I. Introduction

As the complexity of deep neural networks (DNNs) grows, we must design manycore-based accelerators with multiple processing elements (PEs) on a single chip. Three-dimensional (3D) integration and 2.5D interposers are two enabling technologies that enable high degrees of integration to design suitable manycore architectures for accelerating deep DNN workloads. The processing elements (PEs) in both 2.5D and 3D architectures need to be connected via an efficient on-chip communication network to reduce the amount of data movement.

Chiplet-based 2.5D architectures that integrate multiple small dies on an interposer are drawing the attention of leading silicon manufacturers due to their higher energy efficiency and lower fabrication cost than monolithic planar big chips [1]. Chiplet-based systems connect multiple small dies (chiplets) through a network-on-interposer (NoI). Manufacturing several smaller chiplets and combining them into a single system leads to the functionality of a larger chip while maintaining the cost advantages of the smaller chips [2] [3].

Three-dimensional (3D) integration is another technology that enables designers to design high-performance and energy-efficient manycore architectures [4]. Both through-silicon via (TSV)- and emerging monolithic 3D (M3D) enables integration of multiple PEs in a single system. However, the achievable performance of conventional TSV-based 3D systems is ultimately bottlenecked by the horizontal wires (wires in each planar die). Hence, TSV-based architectures do not realize the full potential of 3D integration. M3D integration opens up the possibility of integrating PEs using multiple layers by utilizing nano-scale monolithic inter-tier vias (MIVs), reducing the effective wire length. This leads to better performance and energy efficiency. In addition, M3D provides better heat dissipation than TSV-based designs. Due to better thermal conductivity and extremely thin inter-layer dielectric (ILD), heat is easily dissipated, reducing thermal hotspots [5].

Integrating several PEs or chiplets in a single system introduces additional data exchange. This necessitates the design and optimization of the interconnection network, which is the communication backbone of the 2.5D/3D system. This on-chip data exchange is exacerbated specifically for emerging machine learning (ML) applications. Deep neural networks such as convolutional neural networks (CNNs), Graph Neural Networks (GNNs), Transformer models, and their variants are employed in a range of applications, including autonomous vehicles, machine translation, video analytics, recommendation systems, and social networks [6] [7] [8]. All these ML applications give rise to unique on-chip traffic patterns when mapped onto a manycore system. Hence, designing dataflow-aware manycore accelerators is extremely important.

Recent works have proposed several interconnect architectures for efficient communication between multiple chiplets/PEs on a 2.5D/3D system for ML workloads [9]. Existing NoI/NoC architectures assume a single and typically fixed application workload executed one at a time so that the NoI/NoC can be optimized for a specific application class mapped onto the manycore system. Offline application-specific NoI/NoC optimization is challenging in real-world settings for two main reasons. First, multiple application workloads with varying inputs may need to be executed simultaneously in a real-world scenario (e.g., inferencing for different images or English sentences with varying lengths using the same deep models such as transformers). Second, various workloads may appear simultaneously (e.g., inferencing tasks with different deep models). Specifically, mapping the DNN neural layers onto the chiplets/PEs needs special attention. For example, each neural layer in DNNs typically sends data from layer $L_i$ to layer $L_{i+1}$ (i.e., the data flow graph is mostly linear). Hence, the consecutive neural layers need to be mapped to neighboring chiplets or PEs to reduce latency and improve energy efficiency. This dataflow awareness during the design process is imperative so that the communicating neural layers are highly likely to run on neighboring chiplets/PEs without introducing a significant volume of long-range and multi-hop data exchange. Moreover, in the case of ML algorithms with varying computational kernels (like in the case of transformers,

[1] This work was supported in parts by the National Science Foundation (NSF) Under grants CSR-2308530 and CSR-1955353

influence maximization over graphs), a dataflow-aware strategy can be used for improving energy efficiency and performance. As an example, the fully-connected (FC) layers in Transformers must be mapped in a contiguous manner on the physical NoI/NoC layer to reduce the communication overhead [10]. This is identical to the DNN dataflow.

In this paper, we present design principles of dataflow-aware NoI/NoC architecture specifically targeted towards ML workloads using 2.5D/3D integration. We also highlight the thermal challenges while designing the dataflow-aware manycore architecture.

## II. DATAFLOW-AWARE NoI ARCHITECTURE

2.5D-based manycore systems offer a promising alternative to monolithic chips [1] [3]. Novel 2.5D chiplet platforms provide a new avenue for compact and high-yield architectures for executing various emerging compute- and data-intensive workloads. However, scalable communication between chiplets is particularly challenging due to relatively large physical distances between chiplets, poor technology scaling of electrical wires, and shrinking power budgets. The aforementioned challenges make it difficult to design a viable NoI that can support ultra-high bandwidth, energy-efficient, and low-latency inter-chiplet data transfer without increasing fabrication costs. The demands on the NoI infrastructure will only be exacerbated as application complexity and computational requirements continue to scale. For example, the NoI area overhead alone can be up to 85% of the total system area for a 2.5D-based system [2] [11]. In this section, we present the fundamental idea behind the dataflow-awareness in a chiplet-based system and present a comparative performance evaluation of various NoIs proposed in the literature.

Both application-specific and general-purpose 2.5D chiplet architectures have been explored. Most of these architectures are based on conventional multi-hop interconnection networks, such as mesh or torus [11] [3] [12] [9]. IntAct is one of the earliest architectures demonstrating low latency interconnects on a chiplet-based system using an active interposer [13]. It is a 6-chiplet 96-core architecture with routing logic and peripheral test and programming circuitry like JTAG implemented within the interposer, with a 2D-Mesh interconnection architecture. The SIAM framework enables fast design space exploration of 2.5D-based systems [11]. SIMBA introduces tiling optimizations on fixed 2D-Mesh NoI topology for executing deep models such as ResNet50 [12]. The Kite family of topologies, which are primarily Torus-based, have been proposed for 2.5D-based systems [6]. Recent work discusses the advantages of integrating heterogeneous chiplets on the interposer to reduce design costs [3]. Recently, a compact-packing high-fan-out chiplet-based interconnection architecture called HexaMesh has been proposed [14]. HexaMesh improves bisection bandwidth and has inherently large routers with star-like connections to its nearest possible neighbors. A high-performance and energy-efficient NoI architecture called SWAP was proposed for designing chiplet-based systems for server-scale scenarios, running multiple DNN workloads in parallel [2]. We note that all the above-mentioned NoI architectures (mesh, torus, concentrated mesh, application-specific) principally utilize multi-hop networks, which do not scale with more chiplets. Moreover, these multi-hop NoI architectures create performance bottlenecks for datacenter scale applications, involving multiple concurrent tasks. Recently, a space-filling-curve (SFC) based NoI architecture for 2.5D systems called Floret has been proposed [8]. Fig. 1 shows the SFC-based architecture with six petals connected by a hierarchical top-level network. Floret accelerates datacenter scale ML workloads by employing a dataflow-aware mapping along the chiplets placed in multiple petals of Floret. Moreover, it is scalable for various DNN workloads with similar dataflow.

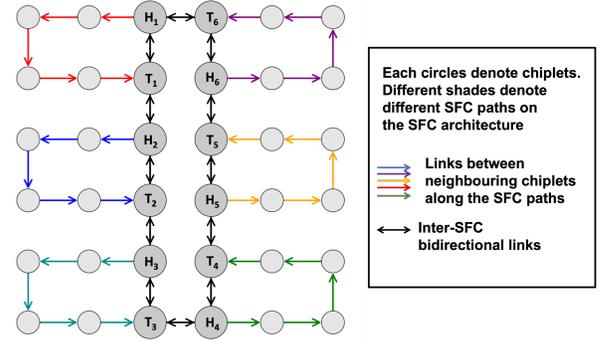

Fig. 1: Illustration of the SFC-based architecture for a 36-chiplet system.

In theory, this dataflow-aware design problem can be viewed as one of embedding a linear ordering (i.e., an SFC) of chiplets over the given topology [15] [16]. However, multiple DNN tasks may need to be dynamically mapped to the system, and each task may consist of different numbers of neural layers. Furthermore, the number of chiplets needed to execute each layer may also vary. Therefore, the problem becomes one of generating *multiple SFCs*, each with its own sequence of chiplets to map the neural layers of any of the tasks [17]. Moreover, as the different DNN tasks complete, the chiplets used for that task need to be *reassigned* to newer tasks. If a consecutive sequence of chiplets is insufficient to accommodate all the layers of a DNN task, the spill over layers will need to utilize chiplets in *other parts* of the NoI (i.e., from other SFCs) to ensure successful completion. During the mapping phase, since the same DNN task may use chiplets from two or more SFCs, it is important to reduce the average number of hops between the tail of an SFC and the head of the next SFC. Therefore, it is imperative to minimize this average path length $d$ between the tail of one SFC to the heads of the other non-overlapping SFCs:

$$\text{Minimize: } d = \frac{1}{p} \sum_{i,j \in [0, \lambda-1]} |t_i - h_j|_{where\ i \neq j,\ p=2\binom{\lambda}{2}} \quad (1)$$

Where $\lambda$ is the number of SFCs and $(h_i, t_i)$ stand for the head and tail of the $i^{th}$ SFC. Here, the distance between any tail-to-head pair is calculated as the Manhattan distance over the 2D grid. Minimizing this average distance measure $d$ is imperative as communication delays between the tail of one SFC and the head of the next SFC can significantly impact the overall system performance. Therefore, the placement of the SFCs and the resulting separation between them in terms of hops become necessary measures to reduce DNN task execution times. Taken together, these factors – i.e., the need to accommodate multiple SFCs, the dynamic nature of mapping multiple DNN tasks to those SFCs, and the need to potentially hop from one SFC to another (for the same task) – all make this a challenging problem, one where classical SFC designs may not apply.

TABLE I: DNN INFERENCE WORKLOADS ALONG WITH THEIR NUMBER OF TRAINABLE PARAMETERS

| Name | DNN model | Dataset | Parameters (in Millions) |
|---|---|---|---|
| $NN_1$ | ResNet18 | Imagenet | 24.76 |
| $NN_2$ | ResNet34 | | 36.5 |
| $NN_3$ | ResNet50 | | 25.94 |
| $NN_4$ | ResNet101 | | 9.42 |
| $NN_5$ | ResNet110 | | 43.6 |
| $NN_6$ | ResNet152 | | 54.84 |
| $NN_7$ | VGG19 | | 93.4 |
| $NN_8$ | DenseNet169 | | 54.84 |
| $NN_9$ | ResNet18 | CIFAR-10 | 11.22 |
| $NN_{10}$ | ResNet34 | | 21.34 |
| $NN_{11}$ | VGG11 | | 9.62 |
| $NN_{12}$ | VGG19 | | 20.42 |
| $NN_{13}$ | GoogLeNet | | 6.16 |

Floret curve is equipped to address all the aforementioned challenges. In particular, Floret connects the chiplets (in the order the neural layers are mapped) along the contiguous path in a two-dimensional (2D) space, as illustrated in Fig. 1. The intuition behind the Floret architecture is to subdivide a multi-dimensional space into smaller contiguous segments (or individual SFCs), and then to stitch those pieces together. Specifically, we leverage the space-filling property to generate a path where a single curve, without any gaps or breaks, traverses the area of the interposer with no closed loops. We first divide the chiplet-based system into multiple SFCs. Each SFC stitches a set of chiplets along the 2D planar path, as illustrated in Fig. 1. Each SFC consists of a head ($H_1, H_2, ..., H_\lambda$) and a tail ($T_1, T_2, ..., T_\lambda$ with $\lambda = 6$ in Fig. 1) connecting a group of chiplets in a contiguous path. We also need to minimize the inter-SFC path length among the non-overlapping SFCs to reduce latency in long-range inter-SFC data exchanges. The advantages of the proposed mapping along the space-filling path of the NoI are two-fold. First, neural layers get mapped to contiguous chiplets and executed in the order they appear until the system is fully utilized. Second, the space-filling NoI architecture, which minimizes the inter-SFC data exchange, reduces the latency when we need to find contiguous chiplet resources belonging to different SFCs. Instead of one monolithic SFC, we use multiple SFCs to introduce inherent redundancy in the system. Even though the Floret curve design is presented for a 2D grid system of chiplets, the design methodology is generic to be extended in principle to other symmetric topologies – e.g., Kite, Butter Donut, Double Butterfly [5] [18]. This is because our algorithm to assign the head-tail pairs simply relies on starting at the center of the NoI and radiating outwards iteratively. In the case of DNNs, given that communication primarily relies on neighboring layers, a simple 2D grid topology is sufficient to serve as the breadboard for generating the Floret curve NoI.

SIMBA, IntAct, and SIAM principally are based on 2D Mesh NoI. We consider SIAM to be representative of this group. Kite is a Torus-based NoI that employs skip connections. Application-specific SWAP NoI is an irregular architecture where the chiplets and the associated routers and links are placed per specific design time considerations for a given set of DNN applications. One of the main differences between SIAM, Kite, SWAP, and Floret is the router port configuration. Each NoI architecture consists of inter-chiplet routers and links. Since each architecture has different connectivity, the distribution of the number of router ports and corresponding link lengths vary.

Fig. 2(a) shows the router-port configuration for Kite, SIAM, SWAP, and Floret for a 100 chiplet system. We observe that four-port routers are the most frequent ones with Kite. SIAM with mesh NoI primarily consists of routers with three and four ports. In contrast, SWAP primarily uses two- and three-port routers, where the links are, on average, longer due to the small-world network approach [2]. However, all the routers in Floret except the heads and tails have only two ports. As a result, the total NoI area of Floret is significantly smaller than other architectures. It should be noted that only reducing the number of links and router port size on their own does not necessarily lead to performance and energy efficiency. To achieve these benefits, it is crucial to consider the length of the links between routers, as the communication delay depends on the link lengths. Fig. 2(b) shows the number of links for different architectures. Kite, for example, has mainly two-hop links, and the routers are inherently bigger. SIAM, principally a 2D-Mesh, has single hop link connections to its neighboring chiplets. SWAP has fewer links and smaller router ports, but not all links are necessarily single-hop. SWAP also has some longer links, with four or five hops. Floret mainly consists of routers with fewer ports, and most links are one-hop connections. In the top-level network, we allow the tail of one SFC to communicate with the heads of other SFCs separated by at most three hops. Within each SFC, all the intra-SFC connections are single hops with small router ports. Smaller routers and reduced link lengths (Fig. 2) in Floret reduce NoI energy, area, and the fabrication costs. Next, we discuss the performance-energy-area-fabrication cost trade-offs among different NoI architectures.

We evaluate the NoI architectures by considering a wide range of concurrent DNNs for inferencing. Table I shows different DNNs and the individual number of parameters. Table II shows the list of multiple concurrent DNN workloads for inferencing simultaneously, representing a datacenter-scale scenario. We consider a 2.5D architecture with 100 chiplets for this performance evaluation. In this work, we aim to enable the acceleration of machine learning (ML) applications using 2.5D/3D architectures. Hence, we consider processing-in-memory (PIM)-based chiplets as the computing platforms.

ReRAM-based PIM is the enabling technology to accelerate DNN inferencing. It should be noted that all the architectures and associated design optimization methodologies are also applicable to other crossbar array (CBA)-based PIM chiplets such as SRAM, STT-MRAM, FeFETs, and many different types of chiplets can be adopted too [2]. Note that the DNNs considered in our performance evaluation consist of linear (VGG), residual (ResNet), and dense (DenseNet) connections. Moreover, all the DNNs consist of fully connected and convolution layers. Each layer of the DNN contains a higher order of multi-bit weights (e.g., ResNet-152 on ImageNet with about 54.8M parameters, VGG19 on ImageNet with 93.4M parameters). However, mapping different DNNs dynamically to a chiplet-based system is challenging. The common property of DNN inference tasks is that activations flow from the $i^{th}$ neural layer to the $(i+1)^{th}$ layer. Hence, there is a need to maintain contiguity on the physical NoI configuration, to the extent possible, between any two consecutive neural layers to reduce communication overhead. It should be noted that the neural networks with skip connections (such as U-Net models for image segmentation or ResNet models) will require communication between non-contiguous layers. However, activations exchanged among non-contiguous layers are still limited. For example, in ResNet34, linear activations are 4.5× higher when compared to skip connections, which are about 19% of the total activations propagated in a single pass. In that case, the inter-chiplet data exchange will involve multiple single-hop paths. This, in turn, will degrade the performance and energy efficiency of the NoI.

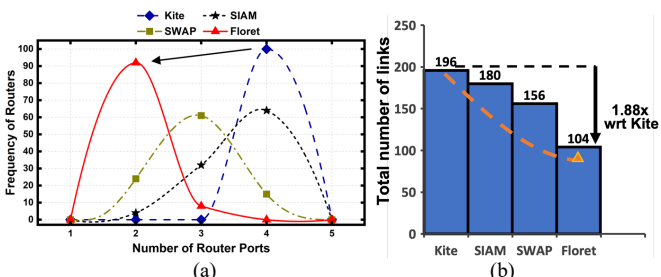

Fig. 2: (a)Variation of router-port configuration for Kite, SIAM, SWAP, and Floret. (b) Comparison of the total number of links for Kite, SIAM, SWAP, and Floret for a 2.5D system with 100 chiplets.

Table II: LIST OF CONCURRENT DNN TASKS ALONG WITH THEIR TOTAL NUMBER OF PARAMETERS FOR 100-CHIPLET SYSTEM (DATASET=IMAGENET)

| Name | DNN model | Total Parameters |
|---|---|---|
| $WL_1$ | $16NN_3 \to NN_8 \to 3NN_6 \to 4NN_5 \to 2NN_1 \to NN_2 \to NN_4$ | 1.1B |
| $WL_2$ | $2NN_6 \to NN_7 \to 7NN_5 \to 4NN_4 \to 2NN_7 \to NN_3 \to NN_1$ | 1.4B |
| $WL_3$ | $12NN_3 \to 9NN_8 \to 3NN_5 \to 10NN_1 \to 12NN_3 \to 5NN_4 \to NN_7$ | 8.8B |
| $WL_4$ | $NN_2 \to 3NN_8 \to 5NN_6 \to 4NN_2 \to 3NN_3 \to 4NN_4 \to 2NN_7$ | 3.8B |
| $WL_5$ | $NN_6 \to 3NN_7 \to 4NN_4 \to 6NN_8 \to 4NN_6 \to 3NN_4 \to 2NN_7$ | 1.8B |

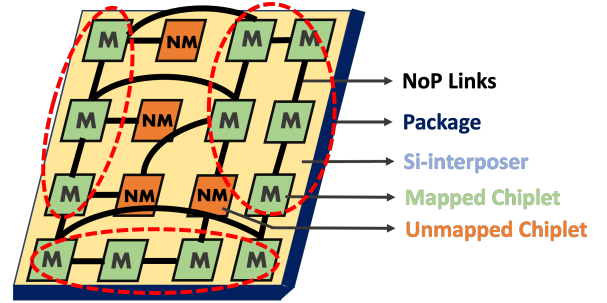

Fig. 4: Illustration of the SWAP architecture for a chiplet-based system with few mapped (M) and unmapped (NM) chiplets. Since the system is optimized at the design time, this leads to poor resource utilization during runtime.

**Performance**: Fig. 3 presents the NoI latency for Floret and the baseline designs (Kite, SIAM, and SWAP) for the 100 chiplet system. Latency is normalized with respect to that of Floret. For instance, we observe that Floret outperforms both the baseline designs (Kite and SIAM) with up to 2.24× improvements in latency. Kite, SIAM, and SWAP incorporate regular NoI topologies and consist of several links that are not necessary for DNN workloads. We map each DNN layer in Kite, SIAM, and SWAP following a greedy mapping strategy. The key idea is to map consecutive DNN layers to chiplets separated by the least number of hops. However, as these three architectures have multi-hop paths between chiplets, finding contiguous available chiplets is not feasible with increasing number of concurrent DNNs. Fig. 4 shows a representative example of SWAP architecture. As we map the neural layers to the chiplets with least communication overhead, this leads to multiple unmapped (NM) chiplets. Most importantly, for bigger system sizes, the multi-hop paths increase even more and give rise to lower resource utilization. On the contrary, Floret always ensures that communicating DNN layers are mapped to contiguous chiplets and utilizes all available resources. The mapping algorithm treats the list of tasks (W) as a queue, assigning one DNN task at a time to avoid deadlock in the case of Floret. Deadlocks could happen only if either there is a cyclic dependency between two tasks (not possible here as DNN tasks are mutually independent), or if there are two concurrent mapping threads that are waiting for one another to release their resources (also not possible due to the sequential queue-based mapping of the DNNs) [8].

**Energy**: By having smaller routers and hence reducing the unnecessary links, Floret not only decreases the inference latency but improves energy efficiency. The energy consumption improvements compared to Kite, SIAM, and SWAP are shown in Fig. 5 for the 100 chiplet system. Energy consumption is normalized with respect to Floret. On average, we observe a 1.65× and 2.8× lower energy than SIAM and Kite, respectively.

**Cost**: NoI comprises around 85% of the total 2.5D system area. Hence, the overall fabrication cost depends on the NoI. The normalized fabrication cost of an NoI is expressed as [11]:

$$C_{NoI} = \frac{L_{ref}}{L} \times e^{-D_0(A_{ref}-A_{NoI})} \quad (2)$$

where $L_{ref}$ is the number of chiplets per wafer in the reference system and $L$ is the number of chiplets per wafer for the system under consideration. The parameter $D_0$ represents the wafer defect density, and $A_{ref}$ is the NoI area of the reference system. We consider a 2.5D system designed by AMD with 864 $mm^2$ interposer area and 64 chiplets as the reference in this work [1]. Using (2), we can compare the fabrication cost of two different NoI architectures. For example, NoI fabrication cost for Floret ($C_{Floret}$) is:

$$C_{Floret} = \frac{L_{ref}}{L} \times e^{-D_0(A_{ref}-A_{Floret})} \quad (3)$$

Similarly, the fabrication cost of the mesh-based SIAM NoI is:

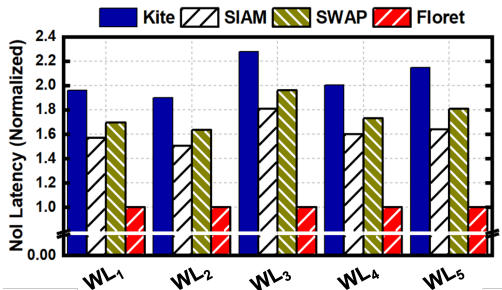

Fig. 3: Comparison of NoI latency for 2.5D system 100 chiplets.

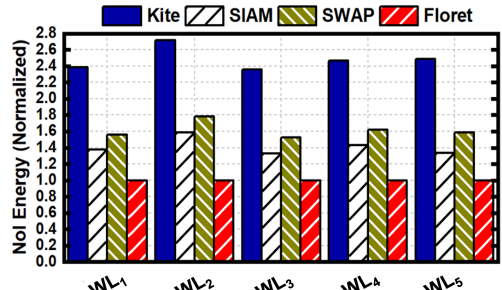

Fig. 5: Comparison of NoI energy for 2.5D system 100 chiplets.

$$C_{SIAM} = \frac{L_{ref}}{L} \times e^{-D_0(A_{ref} - A_{SIAM})} \quad (4)$$

where $A_{Floret}$ and $A_{SIAM}$ correspond to the total NoI area of Floret and SIAM, respectively. Therefore, the fabrication cost of Floret with respect to SIAM can be expressed as:

$$\frac{C_{Floret}}{C_{SIAM}} = e^{-D_0(A_{SIAM} - A_{Floret})} \quad (5)$$

The relative fabrication cost of Floret and other architectures like SIAM principally boils down to the difference between the two NoI areas (5). Since the NoI area increases with increasing number of router ports and NoI links, the corresponding fabrication cost also increases. For instance, Floret reduces fabrication cost by about 2.8×, 2.1×, and 1.89× with respect to Kite, SIAM, and SWAP for a 100-chiplet system, respectively. Floret effectively has smaller router ports and associated links, reducing the fabrication costs. As the scale of datacenter applications is expected to reach an order of 100s of TOPS with billions of storage parameters (equivalent to thousands of chiplets), the fabrication cost becomes an essential component for the affordability of such a system [19]. It is crucial to complement the low fabrication cost with performance and energy benefits. In summary, smaller routers and fewer links along the SFC paths enable Floret to achieve lower latency and fabrication costs with higher energy efficiency than any other existing NoI architecture.

## III. DATAFLOW-AWARE 3D NOC ARCHITECTURE

Thermal bottleneck is not a significant concern in 2.5D architectures due to relatively lower power density than an integrated 3D system. 3D architectures are susceptible to high temperature and hence, will affect inference accuracy of the trained model using ReRAM-based PEs. ReRAM-based PEs store the DNN weights and activations as conductance states, which vary with temperature [20]. As temperature increases beyond $330K$, the conductance range (the gap between $G_{ON}$ and $G_{OFF}$) reduces exponentially. As a result, ReRAM crossbar's output can be misinterpreted, leading to poor inference accuracy [20]. Therefore, it is crucial to consider thermal impact in 3D NoC architectures. Prior work addressed the thermal effects on the inference accuracy of ReRAM-based PIM accelerators via weight remapping, weight reordering, row adjustment, error compensation, etc. [20]. These techniques incur performance loss due to delays introduced by additional peripheral circuits to perform weight splitting, reordering, and error compensation. Further, they consider the effect of temperature on achievable accuracy at the crossbar array level. However, as emerging DNNs workloads use hundreds of millions of parameters, we rely on large-scale integration of PEs. In such an integrated system, inter-PE communication constitutes a significant portion (about 30 to 75%) of the overall execution time of these workloads [4].

A Floret-inspired 3D SFC-enabled NoC architecture connects the PEs in the order the neural layers are mapped along the contiguous path formed by the SFC to achieve high performance. Since a highly integrated single-chip 3D structure has more stringent thermal constraints than a 2.5D system, in addition to performance, neural layer mapping should also consider DNN inference accuracy. PEs executing the initial neural layers typically consume more power as they process more activations. Hence, even along the SFC, we should avoid placing too many power-hungry cores along one specific vertical column of the 3D architecture and away from the heat sink to reduce thermal hotspots. The location of the head/tail of each SFC, the number of SFCs and their respective lengths, and the mapping of consecutive neural layers along the SFC need to be determined by solving a multi-objective optimization (MOO) problem to achieve high performance without sacrificing DNN inference accuracy under thermal impact.

Next, we present a comparative performance evaluation of the Floret-enabled NoC design and joint performance-thermal optimized NoC in terms of energy-delay-product (EDP), peak temperature, and the impact on DNN inference accuracy due to thermal noise. For this evaluation, we consider five DNNs $NN_9$-$NN_{13}$ (shown in Table 1) for brevity. In Figs. 6(a)-(c), we observe that Floret-enabled NoC has a higher EDP reduction by 9% on average since it is optimized for performance only. However, sole-performance optimized mapping leads to higher peak temperature in Floret-enabled NoC by $13K$ on average. As a result, thermal noise and reduced conductance range degrade the DNN inference accuracy in Floret-enabled NoC by up to 11%. Figs. 7(a) and (b) show the thermal hotspot results for the bottom tier (farthest from the heat sink) for Floret-enabled NoC and joint performance-thermal optimized NoC design running $NN_{10}$ (ResNet34) on a 100 PE system as an example. With only performance-optimized neural layer to PE mapping, i.e., the Floret-enabled NoC design, we observe $17K$ higher peak temperature and more thermal hotspots compared to a joint performance-thermal optimized design. This highlights the merit of jointly optimizing performance and temperature objectives for executing DNNs.

## IV. UNIQUE CHALLENGES WITH DATAFLOW AWARE DESIGN

So far, we have discussed how the dataflow-aware design is essential for designing hardware architectures required for training/inferencing with various DNN models. However, for emerging ML workloads such as Transformer models, the awareness needs to be augmented to address complex data movement, memory hierarchy, and latency challenges [10]. Each Encoder block within a Transformer consists of two major functional modules: multi-head self-attention and feed-forward (FF) [21] [22]. Following these two functional modules, there is a residual block to add the input and the output and to perform the layer norm operation. The multi-head attention layer receives data from the input embedding or previous encoder block. The decoder stack also consists of $k$ identical blocks with an extra cross-attention layer to connect with the output from the last encoder stack. For every self-attention layer within a Transformer encoder, all the compute units must be rewritten before the matrix-vector multiplication (MVM) operation is performed since the inputs change dynamically for each token and must be stored in internal registers or on-chip buffers. Traditionally, crossbar-based PIM platforms can directly be employed for the MVM operations involved in DNN workloads. However, such intermediate matrices will require substantial storage capacity or frequent updates in this case. Hence, traditional nonvolatile memory (NVM)-based PIM architectures are unsuitable here due to their limited write endurance. For example, for BERT-Base and Bert-Tiny, intermediate matrices take up to 8.98× and 2.06× of original weight matrix storage, respectively. This storage will grow even more for bigger models, which cannot be stored by using more resources and by remaining within the reticle limit of the 2.5D system with an acceptable yield. Moreover, due to various types of computational kernels involved in Transformer models, we require different types of processing elements, such as Tensor cores, GPUs, DRAMs, and proces sing-in-memory (PIM)-based accelerators on the same system. The FF network consists of two consecutive FC layers, which are large static hidden layers. Like DNN models, the fully connected layers have fixed sizes and sparse weight updates compared to encoder outputs. Data always flows from the $i^{th}$ to the $(i+1)^{th}$ chiplet. Hence, contiguity should be maintained on the physical NoI layer, to the extent possible, between any two consecutive chiplets to reduce the communication overhead. Therefore, we can connect the ReRAM chiplets/PEs using space-filling curves (SFCs). However, the end-to-end transformer model exhibits significant heterogeneity in its computational kernels, necessitating the integration of different types of hardware modules on a single system for high-performance and energy-efficient acceleration. The dataflow-aware NoI/NoC for a part of the computational kernel could be created as a hardware macro with an SFC-based architecture. The other hardware modules must be suitably integrated with this dataflow-aware hardware macro. This is a challenging design space that needs to be thoroughly explored.

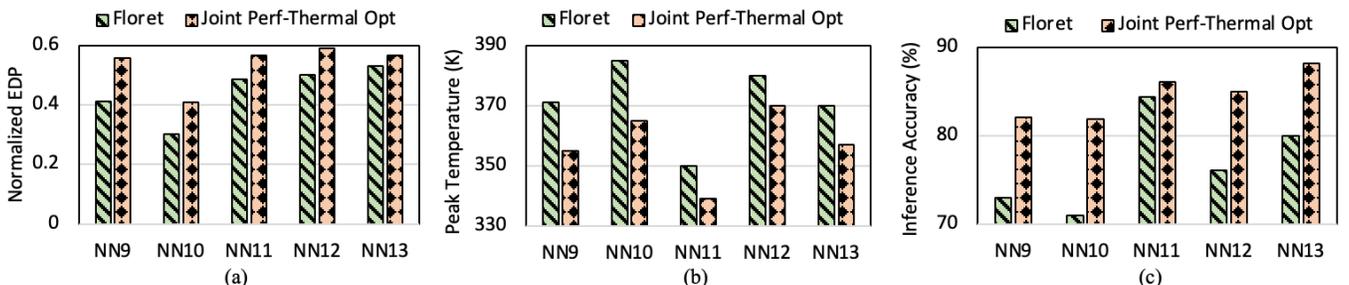

Fig. 6: (a) EDP comparison; (b) Peak temperature comparison; (c) Impact on DNN inference accuracy due to thermal noise on 100 PE NoC-enabled 3D system.

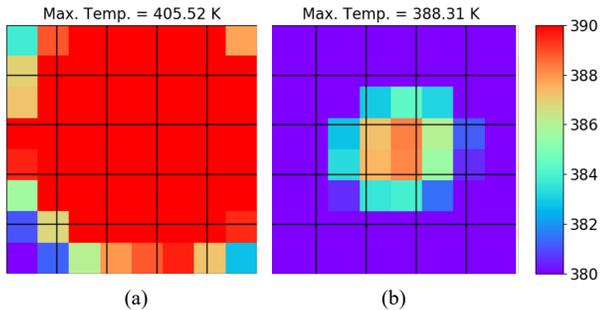

(a)    (b)

Fig. 7: Thermal hotspots in the bottom tier for $NN_{10}$ (ResNet34) running on 100 PE system with (a) Floret-based 3D-NoC and (b) Thermal-aware 3D-NoC.

## V. Conclusion

Datacenters require significant compute and storage resources. Manycore architectures enabled by emerging 2.5D/3D integration technology are enablers for achieving datacenter-scale performance with small form factor designs. The achievable performance and energy efficiency of the manycore architectures depends on the on-chip communication infrastructure (NoI/NoC). NoI/NoC architectures can be optimized by incorporating dataflow awareness inherent in various ML applications. In this paper, we highlight the design principles of a dataflow-aware manycore platform tailor-made for various types of ML workloads. We consider the design challenges with both 2.5D interposer- and 3D integration-enabled architectures. We also highlight future research directions that need to be pursued to leverage the benefits introduced by the dataflow-aware design.